# Analytical, numerical and experimental investigation of a tunable, nonlinear multi-degree-of-freedom parametrically excited amplifier

A. Dolev[1], I. Bucher[1]
[1] Dynamics Laboratory, Mechanical Engineering, Technion
Technion City, Haifa 3200003, Israel
e-mail: **amitdtechnion@gmail.com**

## Abstract
A tunable, multi-degree-of-freedom, parametrically excited amplifier is introduced as an apparatus capable of shifting slow, weak signals to higher frequencies, by exploiting the amplifier natural resonances via controlled parametric excitation and nonlinear feedback. This device can find use as a signal amplifier and as a spectrum control device. A tuned dual-frequency signal is created to parametrically excite (pump) the system and produce the desired energy shift. The pump signal is applied by a controlled electromechanical actuator, and is modified in-situ according to the slow frequency signal and the desired response. A three DOF model is introduced, and the governing nonlinear EOM are derived and solved analytically via asymptotic methods and verified with numerical simulations. A parametric design of an experimental rig was carried out, and the importance of experimental calibration and system identification is emphasized. Some preliminary experimental results are provided with favorable agreement with the theory.

## 1    Introduction

The phenomenon of parametric resonance [1,2] in mechanical systems has been studied extensively in the past [3–5], and in recent times has gained renewed interest by researchers, due to technological advances especially in the field of nano- and micro-electro mechanical systems. Parametric resonance is utilized in mechanical system to create parametric resonators, which are used in sensing [6,7], actuation [8] and filtering [9] applications. These resonators are superior to linear ones in terms of amplification and sensitivity. While the parametric amplifier response is bounded by nonlinearities [10], the time-invariant realizations are bounded by nonlinear stiffness and damping [11]. Although the parametric resonance has found use in some mechanical applications, it may also have devastating effects in some cases, such in the case of ship instability [12] or sheet metal coating [13], therefore should be avoided in many instances.

Parametric resonators can be roughly divided according to their operating mode. Degenerate resonators are parametrically excited (pumped), by a frequency that equals to twice the response frequency, usually close to twice the natural frequency. On the other hand, non-degenerate resonators are not bound to be pumped according to the previously mentioned ratio [14], and some exploit a combination of frequencies, making them more versatile and potentially tunable.

The present work extends a previous work [15] in the field of tunable, nonlinear parametrically excited amplifiers, from a single-degree-of-freedom (SDOF) to multi-degree-of-freedom (MDOF) amplifiers. In the previous work, two operating modes, degenerate and non-degenerate, were utilized simultaneously to drive an electromechanical actuator forming a new operating mode. In this mode, the amplifier is pumped with a dual-frequency signal, thus creating a dual-frequency parametric amplifier (DFPA). The DFPA exploits the advantages of both operating modes, large response amplitude and tunability, making it suitable for a wide band of input signals whose frequency is much lower than the DFPA natural frequency. Exploiting the amplifier resonance, energy is transferred from a slow and weak input to a signal with a larger amplitude and a higher frequency, close to the natural frequency of the amplifier. Extension of the work to a MDOF

DFPA allows more flexibility and tunability, at the expanse of increased complexity. Energy can be transferred to each of the DFPA natural frequencies, and the projection of the external force on the natural mode being excited can be evaluated.

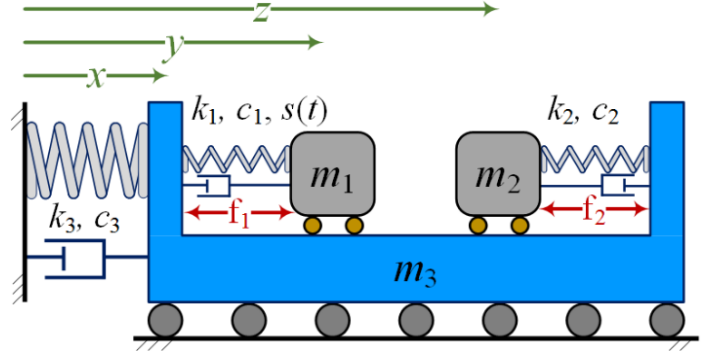

Figure 1: A lightly damped MDOF system with a tunable, nonlinear and time varying stiffness $s(t)$.

In the first section of this paper, analytical model of nonlinear MDOF parametric amplifiers and their response is formulated alongside a brief mathematical explanation. Then, numerical simulations are used to verify the proposed analytical model. Next, the experimental set-up is introduced, flowed by a brief explanation of the system calibration and identification procedure. Afterwards, preliminary experimental results are provided and compared to the analytical solution and numerical simulations. Finally, insights and some small discrepancies between the results are discussed.

## 2 Mathematical derivation and solutions of the EOM

In this section, the model of a MDOF DFPA is introduced alongside the governing equations of motion (EOM). First, the dimensional nonlinear EOM are derived using Hamilton's principle. Then, using several transformations and definition of new parameters, the scaled EOM in modal coordinates are derived, in a form suitable for solution using asymptotic methods. The analytical solution is derived with the method of multiple scales [16].

### 2.1 Derivation of the scaled, nonlinear, modal EOM

In this section, a tunable, nonlinear parametrically excited amplifier having three DOF is investigated as a representative case of a MDOF parametric amplifier. The system's model is depicted in Figure 1, and it is characterized be particle masses $m_\bullet$, linear dashpots $c_\bullet$, linear stiffness $k_\bullet$, and time varying stiffness $s(t)$. Additionally the system is subjected to two external harmonic forces $f_1(t)$ and $f_2(t)$. It proves convenient to define two new coordinates (see Figure 1):

$$u = y - x, \quad v = x - z. \tag{1}$$

Using Hamilton's principle, the EOM of the system in Figure 1 are derived:

$$\begin{pmatrix} \sum m_i & m_1 & -m_2 \\ m_1 & m_1 & 0 \\ -m_2 & 0 & m_2 \end{pmatrix} \begin{Bmatrix} \ddot{x} \\ \ddot{u} \\ \ddot{v} \end{Bmatrix} + \begin{pmatrix} c_3 & 0 & 0 \\ 0 & c_1 & 0 \\ 0 & 0 & c_2 \end{pmatrix} \begin{Bmatrix} \dot{x} \\ \dot{u} \\ \dot{v} \end{Bmatrix} + \begin{pmatrix} k_3 & 0 & 0 \\ 0 & k_p(t) & 0 \\ 0 & 0 & k_2 \end{pmatrix} \begin{Bmatrix} x \\ u \\ v \end{Bmatrix} + \begin{pmatrix} 0 & 0 & 0 \\ 0 & s_3 & 0 \\ 0 & 0 & 0 \end{pmatrix} \begin{Bmatrix} x \\ u \\ v \end{Bmatrix}^3 = \begin{Bmatrix} 0 \\ f_1 \\ f_2 \end{Bmatrix}. \tag{2}$$

Where $s_3$ is a cubic stiffness parameter and $k_p(t)$ is defined as follows:

$$k_p(t) = k_1 \left(1 + \alpha_a \cos(\omega_a t + \varphi_a) + \alpha_b \cos(\omega_b t + \varphi_b)\right). \tag{3}$$

Here two distinct, though algebraically related, pumping frequencies are used $\omega_a$ and $\omega_b$, and the appropriate pumping magnitudes and phases are $\alpha_\bullet$ and $\varphi_\bullet$, $\bullet = a, b$.

It proves convenient to use the undumped linear system normal modes to transform the coordinates, hence the normal modes of the following system were computed:

$$\begin{pmatrix} \sum m_i & m_1 & -m_2 \\ m_1 & m_1 & 0 \\ -m_2 & 0 & m_2 \end{pmatrix} \ddot{\mathbf{q}} + \begin{pmatrix} k_3 & 0 & 0 \\ 0 & k_1 & 0 \\ 0 & 0 & k_2 \end{pmatrix} \mathbf{q} = \mathbf{0}, \qquad \mathbf{q} = \begin{Bmatrix} x \\ u \\ v \end{Bmatrix} \tag{4}$$

The modal matrix, containing the normal modes of the system is defined as follows:

$$\mathbf{\Phi} = \begin{pmatrix} \phi_1 & \phi_2 & \phi_3 \end{pmatrix} \tag{5}$$

Next, the following transformation is used:

$$\mathbf{q} = \varepsilon^\alpha \mathbf{\Phi} \boldsymbol{\eta} \quad \varepsilon \sim O(\hat{\zeta}). \tag{6}$$

Here $\varepsilon$ is some measure of the modal damping and is assumed small:

$$0 < \varepsilon \ll 1, \tag{7}$$

and $\alpha$ is to be determined later, when the problem's various magnitudes' scaling is analyzed.

Substituting Eq.(6) to Eq.(4) and multiplying it by $\varepsilon^{-\alpha} \mathbf{\Phi}^T$ from the left, leads to the following EOM in modal coordinates:

$$\mathbf{I} \ddot{\boldsymbol{\eta}} + 2\hat{\boldsymbol{\zeta}} \boldsymbol{\Omega}_n \dot{\boldsymbol{\eta}} + \boldsymbol{\Omega}_n^2 \boldsymbol{\eta} + \mathbf{\Phi}^T \mathbf{K}(t) \mathbf{\Phi} \boldsymbol{\eta} + \varepsilon^{2\alpha} \mathbf{\Phi}^T \mathbf{S} (\mathbf{\Phi} \boldsymbol{\eta})^3 = \varepsilon^{-\alpha} \mathbf{\Phi}^T \mathbf{Q}$$

$$\hat{\boldsymbol{\zeta}} = \begin{pmatrix} \hat{\zeta}_1 & 0 & 0 \\ 0 & \hat{\zeta}_2 & 0 \\ 0 & 0 & \hat{\zeta}_3 \end{pmatrix}, \quad \boldsymbol{\Omega}_n = \begin{pmatrix} \omega_1 & 0 & 0 \\ 0 & \omega_2 & 0 \\ 0 & 0 & \omega_3 \end{pmatrix}, \quad \mathbf{S} = \begin{pmatrix} 0 & 0 & 0 \\ 0 & s_3 & 0 \\ 0 & 0 & 0 \end{pmatrix}, \tag{8}$$

$$\mathbf{K}(t) = \begin{pmatrix} 0 & 0 & 0 \\ 0 & k_1 \begin{pmatrix} \alpha_a \cos(\omega_a t + \varphi_a) + \\ +\alpha_b \cos(\omega_b t + \varphi_b) \end{pmatrix} & 0 \\ 0 & 0 & 0 \end{pmatrix},$$

Here $\hat{\zeta}_\bullet$ are the modal damping coefficients and $\omega_i, i = 1, 2, 3$ are the natural frequencies.

Dimensionless time is now introduce in the following form:

$$\tau = \hat{\omega} t, \tag{9}$$

where $\hat{\omega}$ is the characteristic frequency of the response, and $\varepsilon^\beta \sim O(\hat{\omega})$, such that:

$$\hat{\omega} = \varepsilon^\beta \tilde{\omega}, \quad \tilde{\omega} \sim O(1). \tag{10}$$

Substituting Eq.(9) and Eq.(10) to Eq.(8) and dividing by $\varepsilon^{2\beta} \tilde{\omega}^2$ leads to the following EOM:

$$\mathbf{I} \boldsymbol{\eta}'' + 2\varepsilon^{-\beta} \hat{\boldsymbol{\zeta}} \hat{\boldsymbol{\chi}} \boldsymbol{\eta}' + \varepsilon^{-2\beta} \hat{\boldsymbol{\chi}}^2 \boldsymbol{\eta} + \varepsilon^{-2\beta} \frac{1}{\tilde{\omega}^2} \mathbf{\Phi}^T \mathbf{K}(\tau) \mathbf{\Phi} \boldsymbol{\eta} + \frac{1}{\tilde{\omega}^2} \varepsilon^{2(\alpha-\beta)} \mathbf{\Phi}^T \mathbf{S} (\mathbf{\Phi} \boldsymbol{\eta})^3 = \frac{1}{\tilde{\omega}^2} \varepsilon^{-(\alpha+2\beta)} \mathbf{\Phi}^T \mathbf{Q}$$

$$\hat{\boldsymbol{\chi}} = \begin{pmatrix} \hat{\chi}_1 & 0 & 0 \\ 0 & \hat{\chi}_2 & 0 \\ 0 & 0 & \hat{\chi}_3 \end{pmatrix}, \quad \hat{\chi}_\bullet = \frac{\omega_\bullet}{\tilde{\omega}} \tag{11}$$

where $\partial/\partial\tau \equiv \bullet'$.

Equation (11) is rewritten in the following form:

$$\mathbf{I}\boldsymbol{\eta}'' + \boldsymbol{\chi}^2\boldsymbol{\eta} = \mathbf{P} - \varepsilon\left(2\zeta\chi\boldsymbol{\eta}' + \mathbf{K}_\gamma(\tau)\boldsymbol{\eta} + \boldsymbol{\kappa}\tilde{\boldsymbol{\eta}}\right). \tag{12}$$

Where light damping, weak pumping and weak nonlinearity are assumed:

$$\begin{aligned}
\hat{\chi} &= \varepsilon^\beta \chi \\
\hat{\zeta}_\bullet &= \varepsilon \zeta_\bullet & &\rightarrow & \hat{\zeta} &= \varepsilon \zeta \\
\alpha_\bullet &= \varepsilon^{1+2\beta} \gamma_\bullet \tilde{\omega}^2 & &\rightarrow & \varepsilon^{-2\beta} \frac{1}{\tilde{\omega}^2}\boldsymbol{\Phi}^T \mathbf{K}(\tau)\boldsymbol{\Phi} &= \varepsilon \mathbf{K}_\gamma(\tau) \\
s_3 &= \varepsilon^{1-2(\alpha-\beta)} \kappa \tilde{\omega}^2 & &\rightarrow & \frac{1}{\tilde{\omega}^2}\varepsilon^{2(\alpha-\beta)} \boldsymbol{\Phi}^T \mathbf{S}(\boldsymbol{\Phi}\boldsymbol{\eta})^3 &= \varepsilon \boldsymbol{\kappa}\tilde{\boldsymbol{\eta}} \\
\mathbf{P} &= \frac{1}{\tilde{\omega}^2}\varepsilon^{-(\alpha+2\beta)}\boldsymbol{\Phi}^T \mathbf{Q}
\end{aligned} \tag{13}$$

the coefficients matrix $\boldsymbol{\kappa}$ and the vector $\tilde{\boldsymbol{\eta}}$ consisting of nonlinear modal coordinates combinations, are provided in Appendix A.

## 2.2 Solution of the scaled, nonlinear, modal EOM

To derive the analytical solution of the governing EOM, an asymptotic method was used, the method of multiple scales, where the following solution form was assumed:

$$\boldsymbol{\eta}(\varepsilon,\tau) = \boldsymbol{\eta}_0(\tau_0,\tau_1) + \varepsilon\boldsymbol{\eta}_1(\tau_0,\tau_1) \tag{14}$$

The latter has two spatial scales $\boldsymbol{\eta}_0$ and $\boldsymbol{\eta}_1$ and two time scales $\tau_i = \varepsilon^i \tau, i = 0,1$. Substituting Eq.(14) into Eq.(12) and collecting terms in the same order of $\varepsilon$ leads to the following ODEs.

$\underline{\varepsilon^0}:$
$$D_0^2 \boldsymbol{\eta}_0 + \boldsymbol{\chi}^2 \boldsymbol{\eta}_0 = \mathbf{P} \tag{15}$$

$\underline{\varepsilon^1}:$
$$D_0^2 \boldsymbol{\eta}_1 + \boldsymbol{\chi}^2 \boldsymbol{\eta}_1 = -\left(2D_0 D_1 \boldsymbol{\eta}_0 + 2\zeta\chi\boldsymbol{\Omega}_n D_0 \boldsymbol{\eta}_0 + \mathbf{K}_\gamma(\tau)\boldsymbol{\eta}_0 + \boldsymbol{\kappa}\tilde{\boldsymbol{\eta}}_0\right) \tag{16}$$

Here the notation $\partial/\partial\tau_\bullet \equiv D_\bullet$ was adopted.

The solution in complex notation of the zeroth order ODEs, Eq.(15), when only the force $f_2(\tau) = F_2 \cos(\chi_r \tau + \varphi_r)$ exist is:

$$\eta_{\bullet 0} = A_i(\tau_1) e^{i\chi_i \tau} + \underbrace{\frac{\varepsilon^{-(\alpha+2\beta)}}{2\tilde{\omega}^2} \frac{\Phi_{3\bullet} F_2}{\chi_i^2 - \chi_r^2}}_{\Lambda_{\bullet 2}} e^{i(\chi_r \tau + \varphi_r)} + CC, \tag{17}$$

where CC stands for the complex conjugate of the preceding terms. To fully compute the zeroth order solution, the amplitude terms, $A_i(\tau_1)$ need to be found, therefore Eq.(17) is substituted into Eq.(16). The right hand side of Eq.(16) can be computed once $\boldsymbol{\eta}_0$ has been substituted and is provided in Appendix A. The solution greatly depends on the pumping and excitation frequencies.

In the following subsection, the response is derived for the case when the system is excited by a slow force $(\omega_r \ll \omega_1)$, while being pumped by a dual frequency signal $(\omega_a \text{ and } \omega_b)$ in a manner that most of the response energy is close to the second natural frequency. The same procedure can be carried out for the remaining natural frequencies, and is omitted here for brevity.

### 2.2.1 Response near the second natural frequency – dual frequency pumping

In order to transfer energy from a frequency well below any natural frequency to the second natural frequency, $f_2$, the following frequency relationships are assumed (based on [15]):

$$\chi_r = \delta\chi_1, \quad \chi_b \approx \chi_2 - \chi_r, \quad \chi_a \approx 2\chi_2, \qquad 0 < \delta < 1. \tag{18}$$

To derive the analytical response the pumping frequencies are defined as:

$$\chi_b + \chi_{r_2} = \chi_2 + \varepsilon\sigma_1 \quad \chi_a = 2\chi_2 + \varepsilon\sigma_2, \tag{19}$$

where $\sigma_1$ and $\sigma_2$ are two algebraically dependent detuning parameters as will be shown. Substituting Eq.(19) to Eq.(16), the secular terms which need to be set to zero in order to derive $A_i$ are:

$$\underline{\chi_1}: \quad i\xi 2\chi_1 A_1' + i2\zeta_1\chi_1^2 A_1 + 3\kappa\Phi_{21}^2 A_1\left(\Phi_{21}^2 A_1\bar{A}_1 + 2\left(H_2^2 + \Phi_{22}^2 A_2\bar{A}_2 + \Phi_{23}^2 A_3\bar{A}_3\right)\right) = 0$$

$$\underline{\chi_2}: \quad i2\chi_2 A_2' + i2\zeta_2\chi_2^2 A_2 + 3\kappa\Phi_{22}^2 A_2\left(\Phi_{22}^2 A_2\bar{A}_2 + 2\left(H_2^2 + \Phi_{21}^2 A_1\bar{A}_1 + \Phi_{23}^2 A_3\bar{A}_3\right)\right) +$$

$$\qquad + \frac{\Phi_{22}k_1}{2}\left(\gamma_b H_2 e^{i(\varphi_b+\varphi_r+\sigma_1\tau_1)} + \gamma_a \Phi_{22}\bar{A}_2 e^{i(\varphi_a+\sigma_2\tau_1)}\right) = 0 \tag{20}$$

$$\underline{\chi_3}: \quad i2\chi_3 A_3' + i2\zeta_3\chi_3^2 A_3 + 3\kappa\Phi_{23}^2 A_3\left(\Phi_{23}^2 A_3\bar{A}_3 + 2\left(H_2^2 + \Phi_{21}^2 A_1\bar{A}_1 + \Phi_{22} A_2\bar{A}_2\right)\right) = 0$$

here, $H_2 = \Lambda_{12}\Phi_{21} + \Lambda_{22}\Phi_{22} + \Lambda_{32}\Phi_{23}$. Transforming $A_\bullet$ to polar form:

$$A_\bullet(\tau_1) = \frac{1}{2}a_\bullet(\tau_1)e^{i\phi_\bullet(\tau_1)}, \tag{21}$$

and separating to real and imaginary equations, leads to:

$$\underline{\chi_1 - \Re}: \quad a_1\phi_1' = \frac{3\kappa\Phi_{21}^2 a_1}{8\chi_1}\left(\Phi_{21}^2 a_1^2 + 2\left(4H_2^2 + \Phi_{22}^2 a_2^2 + \Phi_{23}^2 a_3^2\right)\right)$$

$$\underline{\chi_1 - \Im}: \quad a_1' = -\zeta_1\chi_1 a_1$$

$$\underline{\chi_2 - \Re}: \quad a_2\phi_2' = \frac{3\kappa\Phi_{22}^2 a_2}{8\chi_2}\left(\Phi_{22}^2 a_2^2 + 2\left(4H_2^2 + \Phi_{21}^2 a_1^2 + \Phi_{23}^2 a_3^2\right)\right) +$$

$$\qquad + \frac{\Phi_{22}k_1}{4\chi_2}\begin{pmatrix} 2\gamma_b H_2\cos(\varphi_b+\varphi_r+\sigma_3\tau_1-\phi_2)+ \\ +\gamma_a\Phi_{22}a_2\cos(\varphi_a+\sigma_4\tau_1-2\phi_2) \end{pmatrix} \tag{22}$$

$$\underline{\chi_2 - \Im}: \quad a_2' = -\zeta_2\chi_2 a_2 - \frac{\Phi_{22}k_1}{4\chi_2}\begin{pmatrix} 2\gamma_b H_2\sin(\varphi_b+\varphi_r+\sigma_3\tau_1-\phi_2)+ \\ +\gamma_a\Phi_{22}a_2\sin(\varphi_a+\sigma_4\tau_1-2\phi_2) \end{pmatrix}$$

$$\underline{\chi_3 - \Re}: \quad a_3\phi_3' = \frac{3\kappa\Phi_{23}^2 a_3}{8\chi_3}\left(\Phi_{23}^2 a_3^2 + 2\left(4H_2^2 + \Phi_{21}^2 a_1^2 + \Phi_{22}^2 a_2^2\right)\right)$$

$$\underline{\chi_3 - \Im}: \quad a_3' = -\zeta_3\chi_3 a_3$$

In order to transform Eq.(22) to an autonomous system, the following functions are defined:

$$\psi_1 = \sigma_1\tau_1 - \phi_2, \quad \psi_2 = \sigma_2\tau_1 - 2\phi_2, \tag{23}$$

and their derivatives are:

$$\psi_1' = \sigma_1 - \phi_2', \quad \psi_2' = \sigma_2 - 2\phi_2'. \tag{24}$$

One is interested in the solutions at steady-state, thus sets $a_\bullet' = \psi_1' = \psi_2' = 0$, and from Eq.(24):

$$\phi_2' = \sigma_1, \quad \phi_2' = \frac{1}{2}\sigma_2. \tag{25}$$

Therefore, one defines::

$$\sigma \triangleq \sigma_1 = \frac{1}{2}\sigma_2, \quad \psi \triangleq \psi_1 = \frac{1}{2}\psi_2 \tag{26}$$

Implementation of Eq.(26) leads to the algebraic relation between the pumping frequencies, which was mentioned previously:

$$\chi_a = 2(\chi_r + \chi_b) \implies \omega_a = 2(\omega_r + \omega_b) \tag{27}$$

The solution at steady state is of interest (i.e., $a_{\bullet 0}$, $\phi_{\bullet 0}$ and $\psi_0$), hence $a'_\bullet = \psi' = 0$ are set once Eq.(26) has been substituted into Eq.(22). From the imaginary equations of $\chi_1$ and $\chi_3$ one has $a_{10} = a_{30} = 0$. Therefore, the real equations of $\chi_1$ and $\chi_3$ are fulfilled, and the only equations remained to be solved are:

$$\underline{\chi_2 - \Re}: \quad a_{20}\sigma = \frac{3\kappa\Phi_{22}^2 a_{20}}{8\chi_2}\left(\Phi_{22}^2 a_{20}^2 + 8H_2^2\right) + $$
$$+ \frac{\Phi_{22}k_1}{4\chi_2}\left(2\gamma_b H_2 \cos(\varphi_b + \varphi_r + \psi_0) + \gamma_a \Phi_{22} a_{20} \cos(\varphi_a + 2\psi_0)\right) \tag{28}$$

$$\underline{\chi_2 - \Im}: \quad a_{20} = -\frac{\Phi_{22}k_1}{4\zeta_2\chi_2^2}\left(2\gamma_b H_2 \sin(\varphi_b + \varphi_r + \psi_0) + \gamma_a \Phi_{22} a_{20} \sin(\varphi_a + 2\psi_0)\right)$$

From the imaginary part of the equation, the second mode amplitude can be computed:

$$a_{20} = -\frac{2\Phi_{22}k_1\gamma_b H_2 \sin(\varphi_b + \varphi_r + \psi_0)}{4\zeta_2\chi_2^2 + \Phi_{22}^2 k_1 \gamma_a \sin(\varphi_a + 2\psi_0)} \tag{29}$$

In order to produce large amplitudes (i.e., amplification), the denominator should approach zero, and it is possible only if the pumping magnitude is set according to the following condition:

$$\frac{4\zeta_2\chi_2^2}{k_1\Phi_{22}^2} \leq \gamma_a. \tag{30}$$

The term on the left hand side of Eq.(30) is the linear stability threshold, $\gamma_{\text{LTH}}$, which is an important system parameter that depends on the modal damping, natural frequency, modal shape and modulated stiffness. Substituting Eq.(29) to the real equation leads to the following nonlinear transcendental equation:

$$\frac{6H_2^2 k_1^2 \gamma_b^2 \kappa \Phi_{22}^6 \sin(\varphi_r + \varphi_b + \psi_0)^3}{\chi_2\left(4\zeta_2\chi_2^2 + k_1\gamma_a\Phi_{22}^2 \sin(\varphi_a + 2\psi_0)\right)^3} + \frac{4\left(3H_2^2\kappa\Phi_{22}^2 - \sigma\chi_2\right)\sin(\varphi_r + \varphi_b + \psi_0)}{\chi_2\left(4\zeta_2\chi_2^2 + k_1\gamma_a\Phi_{22}^2 \sin(\varphi_a + 2\psi_0)\right)} +$$
$$+ \frac{-4\zeta_2\chi_2^2 \cos(\varphi_r + \varphi_b + \psi_0) + k_1\gamma_a\Phi_{22}^2 \sin(\varphi_r - \varphi_a + \varphi_b - \psi_0)}{\chi_2\left(4\zeta_2\chi_2^2 + k_1\gamma_a\Phi_{22}^2 \sin(\varphi_a + 2\psi_0)\right)} = 0 \tag{31}$$

from which the phase $\psi_0$ can be computed, hence the amplitude via Eq.(29). Equation (31) can be transformed using trigonometric identities to a 10$^{\text{th}}$ order polynomial [15] and solved with better computational speed and accuracy than iterative zeros search. A closed form solution is not possible, but once $a_{20}$ and $\psi_0$ are computed, the response of the zeroth order is given by:

$$\eta_1 \approx 2\Lambda_{12}\cos(\chi_r\tau + \varphi_r)$$
$$\eta_2 \approx a_{20}\cos((\chi_a/2)\tau - \psi_0) + 2\Lambda_{22}\cos(\chi_r\tau + \varphi_r) . \tag{32}$$
$$\eta_3 \approx 2\Lambda_{32}\cos(\chi_r\tau + \varphi_r)$$

The contribution of the next order terms, especially close to the natural frequency $\chi_2 \approx \chi_a/2$ and at the external force frequency $\chi_r$, can be derived, but is omitted for brevity. The solution with terms of the first order in $\varepsilon$ is:

$$\eta_1 \approx a_{11}\cos\left((\chi_a/2)\tau + \psi_{11}\right) + a_{11r_2}\cos(\chi_r\tau + \psi_{11r})$$
$$\eta_2 \approx a_{20}\cos\left((\chi_a/2)\tau - \psi_{20}\right) + a_{21r}\cos(\chi_r\tau + \psi_{21r}) \quad . \tag{33}$$
$$\eta_3 \approx a_{31}\cos\left((\chi_a/2)\tau + \psi_{31}\right) + a_{31r}\cos(\chi_r\tau + \psi_{31r})$$

From Eq.(31), ten conjugate solutions are possible; implementation of the reverse transformation leads to five possible real solutions, which are either stable or unstable. The stability analysis is carried out in a standard procedure [16], therefore omitted from the paper, but is shown in the presented results.

## 3    Numerical verification of the DFPA response

In order to verify the analytical results, the DFPA parameters should be chosen according to the assumptions made during the solution. Assumptions regarding the relations between the frequencies were made, to ensure that only certain terms stemming from the nonlinearity are indeed secular. Additionally, assumption regarding the magnitude of the pumping $\alpha_a$ were made, which should be larger than the linear stability threshold $(\gamma_{\text{LTH}} \leftrightarrow \alpha_{\text{LTH}})$ and smaller than unity. Values larger than one are unpractical as their meaning is that negative stiffness is produced. Moreover, to avoid negative stiffness when dual pumping is utilized, the sum of the pumping magnitudes should obey $\alpha_a + \alpha_b < 1$. Therefore the following parameters were chosen:

$$\begin{array}{lll} m_1 = 0.5\,[\text{kg}] & m_2 = 0.25\,[\text{kg}] & m_3 = 5.9\,[\text{kg}] \\ k_1 = 2.53\times10^3\,[\text{N/m}] & k_2 = 1.20\times10^3\,[\text{N/m}] & k_3 = 3.65\times10^4\,[\text{N/m}], \\ \hat{\zeta}_1 = 1\% & \hat{\zeta}_2 = 2\% & \hat{\zeta}_3 = 3\% \end{array} \tag{34}$$

leading to the following natural frequencies, modal matrix and linear stability thresholds:

$$\mathbf{f_n} = \begin{Bmatrix} 9.94 \\ 11.15 \\ 14.156 \end{Bmatrix}[\text{Hz}], \quad \mathbf{\Phi} = \begin{bmatrix} 0.2129 & 0.0254 & 0.3515 \\ 0.7147 & 0.8404 & -0.9759 \\ -0.887 & 1.6033 & 0.9041 \end{bmatrix}\left[\frac{1}{\sqrt{\text{kg}}}\right], \quad \boldsymbol{\alpha}_{\text{LTH}} = \begin{Bmatrix} 12.07 \\ 22.00 \\ 39.38 \end{Bmatrix}[\%] \tag{35}$$

Additionally the external force was set as:

$$\mathrm{f}_2 = 1\cos(0.1\omega_1 t) \tag{36}$$

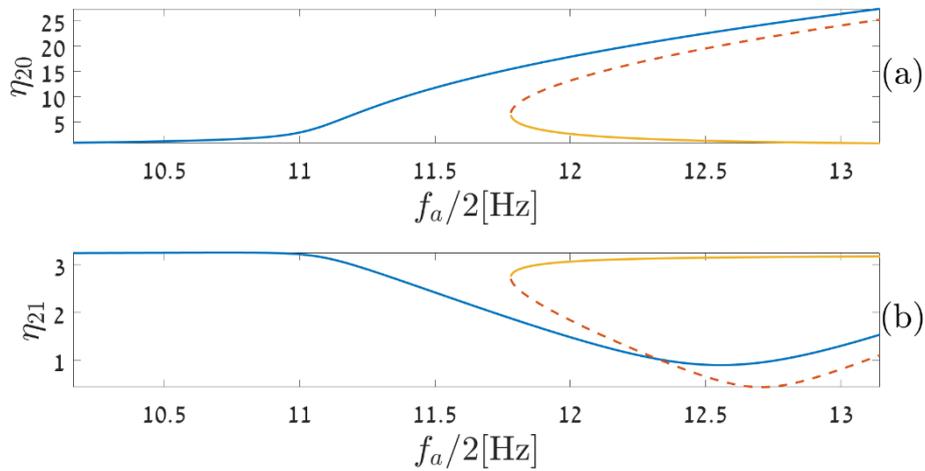

Figure 2: Analytically computed response amplitudes of the second modal coordinate when the system is tuned to shift energy to the second mode. (a) response amplitudes close to the second natural frequency, $f_a/2$, (b) the corresponding response amplitudes at the external force frequency $f_r$. Continuous lines mark stable and dashed lines mark unstable solutions.

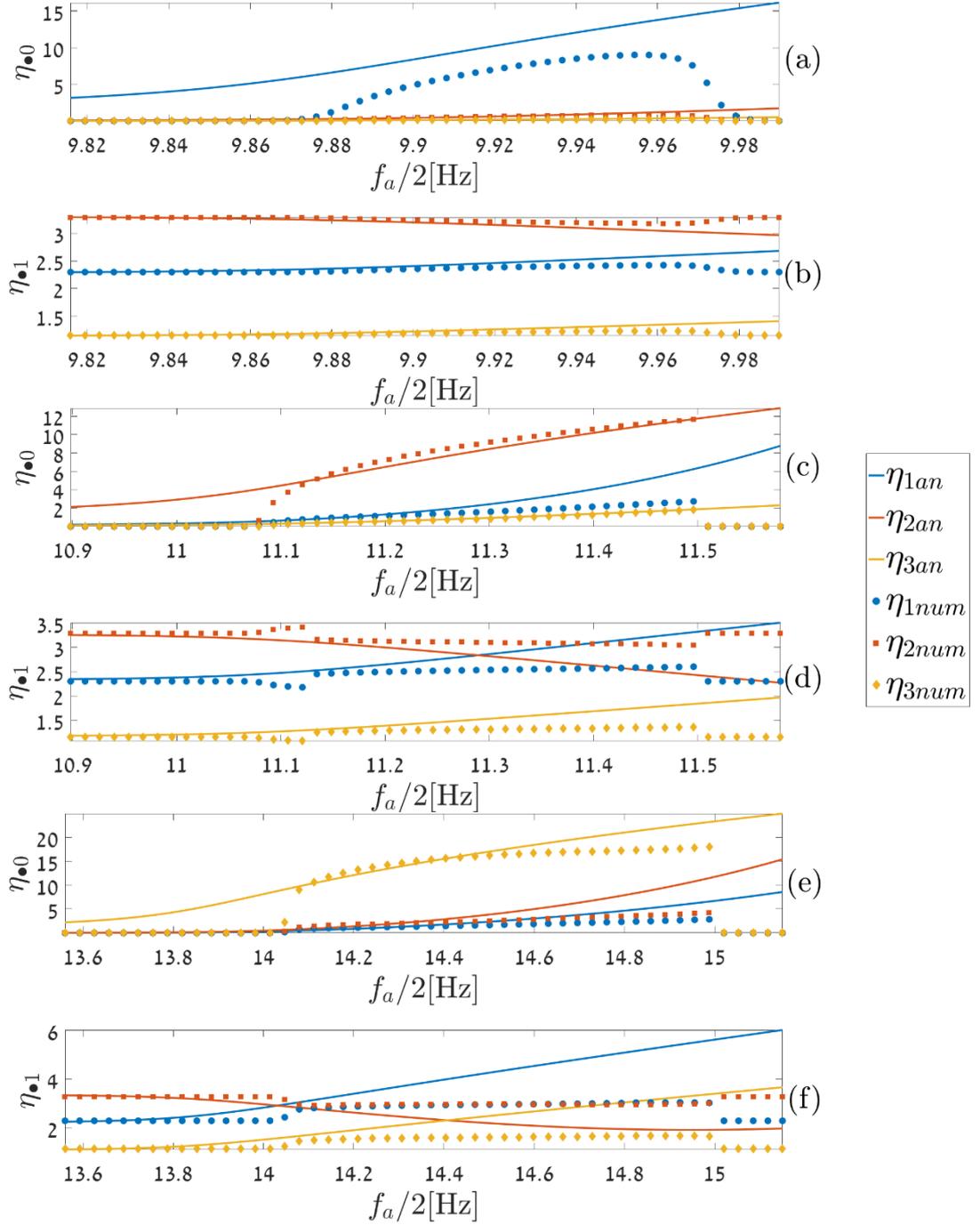

Figure 3: Analytically and numerically computed response amplitudes displayed in modal coordinates, when the system is tuned to shift energy to a specific mode. In the subfigures, only the largest solution branch of the response near the relevant natural frequency is displayed. (a) response amplitudes for $f_a/2 \approx f_1$, (b) corresponding amplitudes at $f_r$, (c) response amplitudes for $f_a/2 \approx f_2$, (d) corresponding amplitudes at $f_r$, (e) response amplitudes for $f_a/2 \approx f_3$, (f) corresponding amplitudes at $f_r$. Analytical solutions are displayed by continuous lines and numerical solutions by data markers (dots, squares and diamonds).

The nonlinearity was set to bring certain amplitudes according to the mode to which energy was transferred. From the problem's various magnitudes' scaling analysis the parameters $\varepsilon, \alpha$ and $\beta$ appearing in Eq.(6), (10) and (13) were set to:

$$\varepsilon = 10^{-2}, \quad \alpha = 2, \quad \beta = -1/2 \tag{37}$$

The analytical solutions of the second modal coordinate $\eta_2$, when the energy is shifted to the second mode are displayed in Figure 2. One can notice that multiple solutions may exist for a given pumping frequency, and the largest amplitude close to the natural frequency, $\sim f_2$, is stable (blue continuous line in Figure 2(a)). For this solution it is noticeable that energy is transferred from the external force frequency, $f_r$, (blue continuous line in Figure 2(b)) to the natural frequency (i.e., as the solution in Figure 2(a) increases the corresponding one in Figure 2(b) decreases). It is clear that this solution is the most significant, as the apparatus is designed to amplify signals and shift energy.

Next, the analytical solutions were verified via comparison to numerical simulations. Notice that only the largest stable amplitudes are displayed in order to avoid confusion. Figure 3 depicts the analytical and numerical responses, when dual-frequency pumping was employed to shift, each time, energy from the external force to a different natural frequency and mode. For each case, the modal amplitudes near the natural frequency and at the external force frequency are shown vs. the pumping frequency. It is noticeable that for all three cases the analytical and numerical solution resemble up to a certain pumping frequency for which the solution is no longer stable, and the numerical solutions drops to zero. Interestingly, for all three cases, as the pumping frequency is increased, more energy is transferred from the second mode to the relevant mode. The comparison verifies the theory according which energy can be transferred to any desired mode by correctly tuned pumping frequencies.

## 4  Experimental set-up, identification, some results and analysis

The experimental system shown in Figure 4 was designed and built to validate the theoretical results and refine the theory as needed. The system consist of an aluminum plate which is suspended by four leaf springs enforcing motion in a single direction as depicted. Two modular masses are attached to the aluminum plate via leaf springs enforcing parallel motion as well. The displacement of the various masses are measured using lased displacement sensors which are attached to the aluminum plate. The forces applied to the masses $m_1$ and $m_2$ are via linear voice coil actuators which are also attached to the aluminum plate. In this configuration the measured coordinates are $x(t)$, $u(t)$ and $v(t)$.

The actuators were driven via a real-time digital signal processor (dSPACE 1104) to allow for easy tuning of the parametric excitation and nonlinear stiffness. In addition, the latter also served as a data question device.

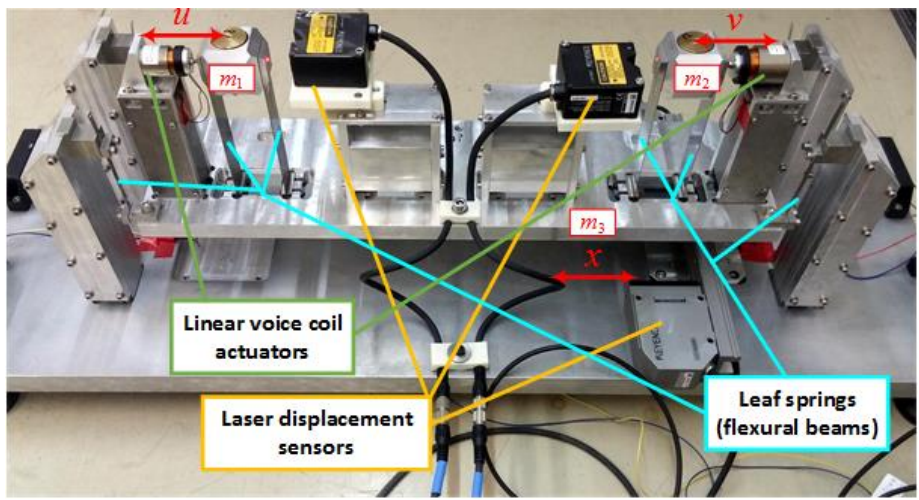

Figure 4: Experimental rig, showing the mechanical parts, the sensors and voice-coil actuators

## 4.1 System calibration and identification

Prior to conducting experiments, the sensors and voice coil actuators must be calibrated, and the system parameters need to be identified. At the beginning, the displacement sensors were calibrated using a manual micrometer. Next, the voice coil actuators were experimentally modeled and their mild static nonlinearity was pre-compensated by adjusting the supplied input signals. This action eliminates the inherent nonlinearity of the applied force created by magnetic imperfection in the voice coils.

Once the system sensors and actuators were calibrated, its parameters were set to be identified. Two frequency responses were produced and measured, where the input forces were $(f_1, f_2) = (1,1)$ and $(f_1, f_2) = (-1,1)$. The natural frequencies, normal modes and damping coefficient were estimated with the aid of Structural Dynamics Toolbox [17] which contains a MIMO identification algorithm. Knowing the system topology, the method of model updating [18] was used to reevaluate the masses and stiffness. The results are provided below:

$$\begin{aligned} & f_1 = 3.031 [\text{Hz}] \quad f_2 = 6.980 [\text{Hz}] \quad f_3 = 8.794 [\text{Hz}] \\ & \hat{\zeta}_1 = 0.5 [\%] \quad \hat{\zeta}_2 = 1.89 [\%] \quad \hat{\zeta}_3 = 2.66 [\%] \\ & m_1 = 0.3501 [\text{kg}] \quad m_2 = 0.3535 [\text{kg}] \quad m_3 = 6.4966 [\text{kg}] \\ & k_1 = 637.4 [\text{N/m}] \quad k_2 = 1009.4 [\text{N/m}] \quad k_3 = 2663 [\text{N/m}] \end{aligned} \qquad (38)$$

and the normalized modes are shown in Figure 5**Error! Reference source not found.**.

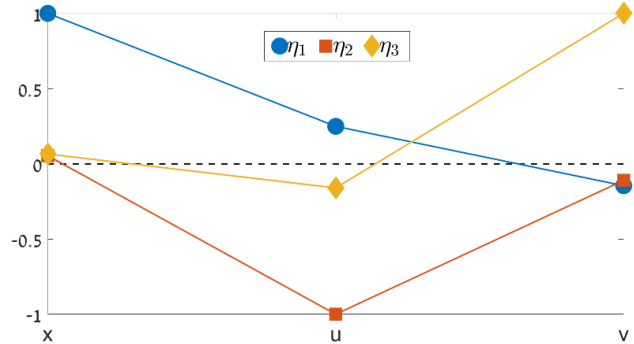

Figure 5: Mode shapes

## 4.2 Principal parametric resonance – verification of linear stability threshold

As explained in Section 2.2.1, the linear stability threshold is an important system parameter and it is proportional to the modal damping. To produce large amplitudes and amplification one needs to know these parameters values (i.e., each mode has a different threshold).

Having identified the linear system parameters and successfully eliminated the inherent actuator static nonlinearities, the model was ready to the stage where it is digitally modified. Next, the parametric excitation and system nonlinearity were added through the actuators as computed in the digital signal processor. The appropriate expressions for pumping and nonlinearity, which appear in the Eq.(2), were realized through a digital feedback, as formulated below.

$$\begin{aligned} & \mathbf{M\ddot{q}} + \mathbf{C\dot{q}} + \mathbf{Kq} = \mathbf{F}_{DE}(t) - \mathbf{F}_{P}(u,t), \quad \mathbf{q} = \{x \quad u \quad v\}^T \\ & \mathbf{F}_{DE} = \cos(\omega_r t + \varphi_r)\{0 \quad 0 \quad f_2\}^T \\ & \mathbf{F}_{P} = k_1 \left( \alpha_a \cos(\omega_a t + \varphi_a) + \alpha_b \cos(\omega_b t + \varphi_b) \right)\{0 \quad u \quad 0\}^T + s_3 \{0 \quad u^3 \quad 0\}^T \end{aligned} \qquad (39)$$

here the matrices $\mathbf{M}$, $\mathbf{C}$ and $\mathbf{K}$ contain the linear system parameters which were estimated in Section 4.1 and provided in Eq.(38). On the right hand side, the forces $\mathbf{F}_{DE}$ and $\mathbf{F}_{P}$ are applied by the actuators fed by the fast digital system, when the direct excitation force applied on $m_2$ and $m_3$ is realized via $\mathbf{F}_{DE}$, and the pumping and nonlinearity are realized via $\mathbf{F}_{P}$. While the former term, $\mathbf{F}_{DE}$, is only time dependent, the latter, $\mathbf{F}_{P}$, is time and position dependent.

The zeroth order solution in $\varepsilon$ of the governing equations for the case of principal parametric resonance [16], when the system is pumped with a single frequency close to twice the natural frequency and without external excitation, is:

$$a_{\bullet 0} = \frac{1}{\Phi_{2\bullet}^2 \sqrt{3\kappa}} \sqrt{4\sigma_\bullet \chi_\bullet \pm 2k_1 \gamma_a \Phi_{2\bullet}^2 \sqrt{1 - \frac{16\zeta_\bullet^2 \chi_\bullet^4}{k_1^2 \gamma_a^2 \Phi_{2\bullet}^4}}}, \quad \chi_a = 2\chi_\bullet + \varepsilon \sigma_\bullet$$
$$\psi_{\bullet 0}^+ = \arcsin\left(\frac{4\zeta_\bullet \chi_\bullet^2}{k_1 \gamma_a \Phi_{2\bullet}^2}\right) + \pi - \varphi_a, \quad \psi_{\bullet 0}^- = -\arcsin\left(\frac{4\zeta_\bullet \chi_\bullet^2}{k_1 \gamma_a \Phi_{2\bullet}^2}\right) - \varphi_a \quad (40)$$

The solution procedure is omitted for brevity as it is similar to the one carried out in Section 2. It is noticeable, that the pumping magnitude $\gamma_a$ must be larger than the linear stability threshold, $\gamma_{LTH} = 4\zeta_\bullet \chi_\bullet^2 / k_1 \Phi_{2\bullet}^2$, for the asymptotic solutions to exist.

The pumping magnitude and nonlinearity were set according to Eq.(40) to bring certain amplitudes when the system is tuned to be excited near the second natural frequency. In order to estimate $\gamma_{LTH}$ for $f_2$ the system was pumped with a magnitude larger than $\gamma_{LTH}$, for which a parametric resonance was produced. Then the pumping magnitude was slowly decreased until no measurable amplitude sustained at steady-state. The pumping magnitude for which no amplitude sustained is denoted $\gamma_{LTH}$. The measured amplitudes in modal coordinate vs. the deviation between the analytical value of $\gamma_{LTH}$ and the applied pumping magnitude, $\gamma$, are shown in Figure 6.

Figure 6(a) depicts the measured amplitude vs. the deviation, when the $\gamma_{LTH}$ was computed according to the estimated $\hat{\zeta}_2$ in Eq.(38). Using these measurements, the actual $\gamma_{LTH}$ was estimated, and the damping $\hat{\zeta}_2$ was updated accordingly, assuming that $k_1$, $\chi_2$ and $\Phi_{22}$ are accurate. A similar experiment was conducted using the updated damping and actual $\gamma_{LTH}$, and the results are shown in Figure 6(b). Comparing the results from the two experiments, the deviation was reduced from about 1.8 to 0.005.

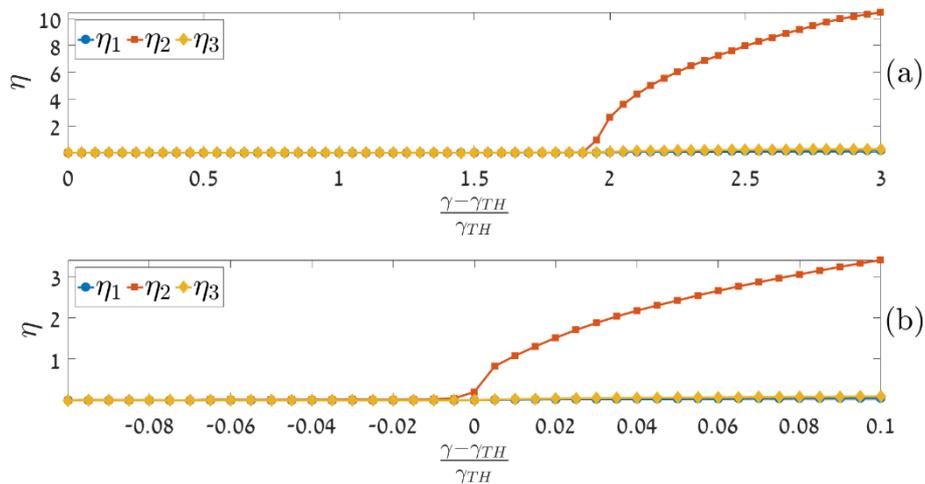

Figure 6: Response amplitude in modal coordinates vs. the deviation of the pumping magnitude from theory before $\hat{\zeta}_2$ was updated (a), and after it and $\gamma_{LTH}$ were updated (b).

## 4.3 Dual-frequency parametric excitation – preliminary experimental results and verification

After the system was calibrated, its parameters identified and a method to derive the relevant linear stability threshold was verified, utilization of the DFPA mode was ready to be use. Based on the analytical solution, Eq.(33), the pumping magnitudes and nonlinearity were chosen to bring certain amplitudes.

The measured frequency response, analytical solution and numerical simulation for the parameters provided in Eq.(41) appear in Figure 7 and Figure 8.

$$s_3 = 4.44 \times 10^7 \left[\text{N/m}^3\right], \quad \alpha_a = 0.24, \quad \alpha_b = 0.24, \quad f_2 = 0.1 [\text{N}]$$
$$f_a \approx 6.98 [\text{Hz}] \quad f_b = 6.677 [\text{Hz}] \quad f_r = 0.303 [\text{Hz}]$$
(41)

The amplifier response, when operated in DFPA mode close to a natural frequency, consists of two main harmonics with a relatively large amplitude in comparison to the other ones. The largest amplitude is at a frequency close to the chosen natural frequency, in this case $f_2$, and an additional amplitude at the external force frequency, $f_r$.

Observing Figure 7, in which the response amplitudes close to $f_2$ are shown, a good resemblance between the numerical simulation and analytical solution can be seen, as seen previously in Figure 3. The experimental results greatly resemble the numerical simulation, although shifted by about $0.3 [\text{Hz}]$. It is important to mention that unlike the simulation, for which the solution dropped to zero at about $7.25 [\text{Hz}]$, the measured amplitude did not. In fact, the measured amplitude reached a value of about $1.35 [\text{mm}]$ which is the allowable value, therefore the experiment did not proceed beyond this frequency. It seems that higher values may have been achieved as the theory predicts.

The amplifier response amplitudes at the external force frequency, $f_r$, are shown in Figure 8, and provide similar insights as Figure 7. The analytical, numerical and experimental results resemble, although the measured response is shifted by about $0.3 [\text{Hz}]$.

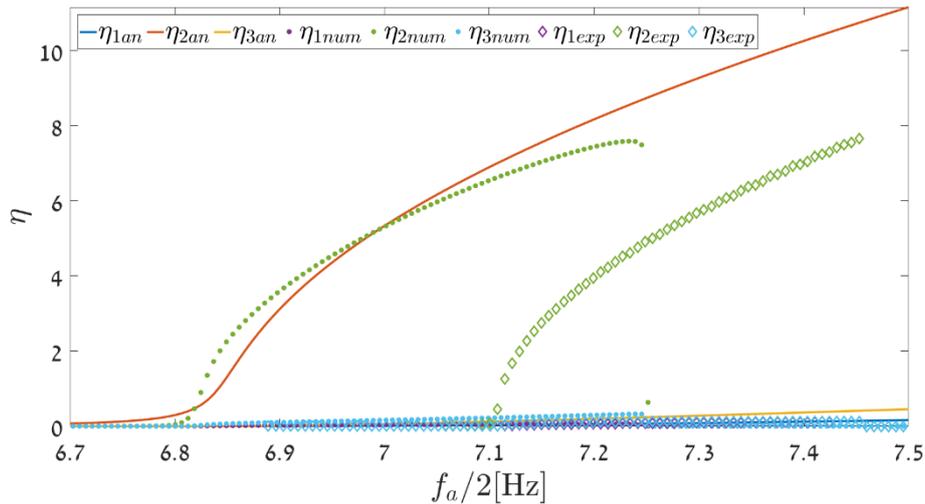

Figure 7: Amplitudes in modal coordinates near the second natural frequency $f_2$ vs. the pumping frequency. Analytical results are shown by continuous lines, numerical results by circular data markers and experimental results by hollow diamond shaped data markers.

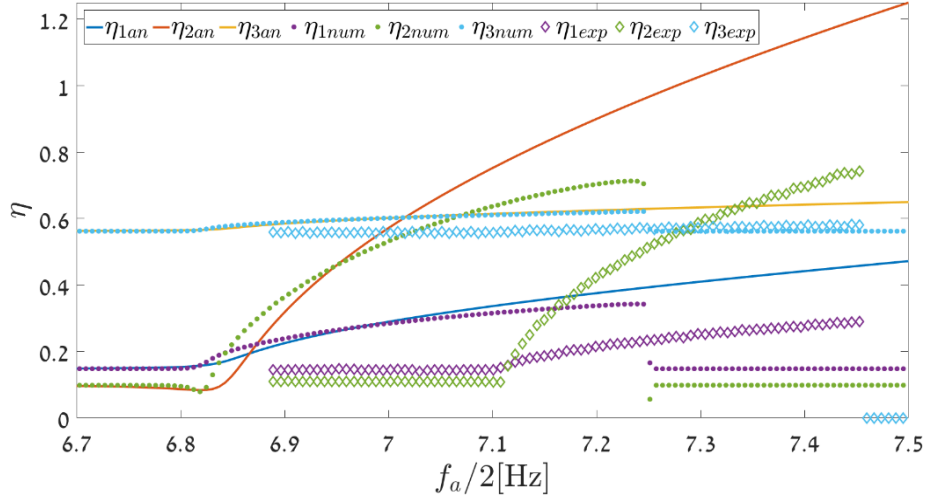

Figure 8: Amplitudes in modal coordinates at the external force frequency $f_r$ vs. the pumping frequency. Analytical results are shown by continuous lines, numerical results by circular data markers and experimental results by hollow diamond shaped data markers.

# 5 Conclusions

A tunable, nonlinear parametrically excited amplifier with three DOF model was introduced. Its governing EOM were derived and solved using the method of multiple scales for the case of combined dual-frequency pumping signal and external excitation. It has been shown that the amplifier can be tuned to shift energy to any of its natural frequencies by a properly tuned pumping signal. These results were verified by numerical simulations shown in Figure 3 and experimental results shown in Section 4.

During the analytical solution an important system parameters were introduced, the linear stability thresholds. In contrast to SDOF systems, for MDOF system these parameters does not solely depend on the damping coefficient, but also on the natural frequencies, modal shapes and modulated stiffness via, $\chi_\bullet$, $\Phi_{2\bullet}$ and $k_1$. Unsurprisingly, it turns out that the topology has a great impact on the ability to parametrically excite a system. Therefore, when designing one care should be taken for the resulting values of the thresholds. Large values of $\alpha_{\text{LTH}}$ exceeding unity result with negative stiffness in practice, which is usually undesired or requires large forces making it energetically inefficient.

An experimental system was built in order to validate the results and refine the theory as needed. Prior to conducting experiments, the system was calibrated and its parameters were estimated. Preliminary results are shown in Section 4 and resemblance between the various solutions was noticed once the linear stability threshold was estimated. According to the estimated $\alpha_{\text{LTH}}$, and assuming no error in the natural frequencies, modal shapes and modulated stiffness, the damping coefficient was reevaluated.

In the nonlinear frequency response shown in Section 4.3, a mild frequency shift was noticed between the measured response and analytical and numerical solutions. Although $0.3 [\text{Hz}]$ shift is small by engineering standards, it means that as expected, there are errors in the estimated system parameters, which are the masses, stiffness and damping. It is relatively easy to notice the error in the estimated natural frequencies, but hard to notice the error in the mode shapes. Nevertheless, the results are satisfactory and validate the model.

# Acknowledgements

This research was supported by the Technion funds for Security research.

# Appendix A

The nonlinear matrix and modal vector are:

$$\boldsymbol{\kappa}^{\mathrm{T}} = \frac{\kappa}{\tilde{\omega}^2} \begin{pmatrix} \Phi_{21}^4 & \Phi_{21}^3\Phi_{22} & \Phi_{21}^3\Phi_{23} \\ 3\Phi_{21}^3\Phi_{22} & 3\Phi_{21}^2\Phi_{22}^2 & 3\Phi_{21}^2\Phi_{22}\Phi_{23} \\ 3\Phi_{21}^2\Phi_{22}^2 & 3\Phi_{21}\Phi_{22}^3 & 3\Phi_{21}\Phi_{22}^2\Phi_{23} \\ \Phi_{21}\Phi_{22}^3 & \Phi_{22}^4 & \Phi_{22}^3\Phi_{23} \\ 3\Phi_{21}^3\Phi_{23} & 3\Phi_{21}^2\Phi_{22}\Phi_{23} & 3\Phi_{21}^2\Phi_{23}^2 \\ 6\Phi_{21}^2\Phi_{22}\Phi_{23} & 6\Phi_{21}\Phi_{22}^2\Phi_{23} & 6\Phi_{21}\Phi_{22}\Phi_{23}^2 \\ 3\Phi_{21}\Phi_{22}^2\Phi_{23} & 3\Phi_{22}^3\Phi_{23} & 3\Phi_{22}^2\Phi_{23}^2 \\ 3\Phi_{21}^2\Phi_{23}^2 & 3\Phi_{21}\Phi_{22}\Phi_{23}^2 & 3\Phi_{21}\Phi_{23}^3 \\ 3\Phi_{21}\Phi_{22}\Phi_{23}^2 & 3\Phi_{22}^2\Phi_{23}^2 & 3\Phi_{22}\Phi_{23}^3 \\ \Phi_{21}\Phi_{23}^3 & \Phi_{22}\Phi_{23}^3 & \Phi_{23}^4 \end{pmatrix}, \quad \tilde{\boldsymbol{\eta}}^{\mathrm{T}} = \begin{Bmatrix} \eta_1^3 \\ \eta_1^2\eta_2 \\ \eta_1\eta_2^2 \\ \eta_2^3 \\ \eta_1^2\eta_3 \\ \eta_1\eta_2\eta_3 \\ \eta_2^2\eta_3 \\ \eta_1\eta_3^2 \\ \eta_2\eta_3^2 \\ \eta_3^3 \end{Bmatrix} \quad (A.1)$$

The right hand side of the Eq.(16) term by term is as follows:

$$2D_0 D_1 \eta_{\bullet 0} = \mathrm{i} 2\chi_\bullet A'_\bullet \mathrm{e}^{\mathrm{i}\chi_\bullet \tau} + \mathrm{CC} \tag{A.2}$$

$$2\zeta_\bullet \chi_\bullet D_0 \eta_{\bullet 0} = 2\zeta_\bullet \chi_\bullet \left( \mathrm{i}\chi_\bullet A_\bullet \mathrm{e}^{\mathrm{i}\chi_\bullet \tau} + \mathrm{i}\chi_r \Lambda_{\bullet 2} \mathrm{e}^{\mathrm{i}(\chi_r \tau + \varphi_r)} \right) + \mathrm{CC} \tag{A.3}$$

$$\left(\mathbf{K}_\gamma(t)\boldsymbol{\eta}_0\right)_\bullet = \frac{\Phi_{2\bullet} k_1}{2} \begin{pmatrix} H_2 \begin{pmatrix} \gamma_a \mathrm{e}^{\mathrm{i}((\chi_a - \chi_r)\tau + \varphi_a - \varphi_r)} + \gamma_a \mathrm{e}^{\mathrm{i}((\chi_a + \chi_r)\tau + \varphi_r + \varphi_a)} + \\ +\gamma_b \mathrm{e}^{\mathrm{i}((\chi_b - \chi_r)\tau + \varphi_b - \varphi_r)} + \gamma_b \mathrm{e}^{\mathrm{i}((\chi_b + \chi_r)\tau + \varphi_b + \varphi_r)} \end{pmatrix} + \\ +\Phi_{21} A_1 \left( \gamma_a \mathrm{e}^{\mathrm{i}((\chi_1 + \chi_a)\tau + \varphi_a)} + \gamma_b \mathrm{e}^{\mathrm{i}((\chi_1 - \chi_b)\tau - \varphi_b)} + \gamma_b \mathrm{e}^{\mathrm{i}((\chi_1 + \chi_b)\tau + \varphi_b)} \right) + \\ +\Phi_{22} A_2 \left( \gamma_a \mathrm{e}^{\mathrm{i}((\chi_2 + \chi_a)\tau + \varphi_a)} + \gamma_b \mathrm{e}^{\mathrm{i}((\chi_2 - \chi_b)\tau - \varphi_b)} + \gamma_b \mathrm{e}^{\mathrm{i}((\chi_2 + \chi_b)\tau + \varphi_b)} \right) + \\ +\Phi_{23} A_3 \left( \gamma_a \mathrm{e}^{\mathrm{i}((\chi_3 + \chi_a)\tau + \varphi_a)} + \gamma_b \mathrm{e}^{\mathrm{i}((\chi_3 - \chi_b)\tau - \varphi_b)} + \gamma_b \mathrm{e}^{\mathrm{i}((\chi_3 + \chi_b)\tau + \varphi_b)} \right) + \\ +\gamma_a \mathrm{e}^{\mathrm{i}((\chi_a - \chi_1)\tau + \varphi_a)} \Phi_{21} \bar{A}_1 + \gamma_a \mathrm{e}^{\mathrm{i}((\chi_a - \chi_2)\tau + \varphi_a)} \Phi_{22} \bar{A}_2 + \gamma_a \mathrm{e}^{\mathrm{i}((\chi_a - \chi_3)\tau + \varphi_a)} \Phi_{23} \bar{A}_3 \end{pmatrix} + \mathrm{CC},$$

(A.4)

where $H_2 = \Lambda_{12}\Phi_{21} + \Lambda_{22}\Phi_{22} + \Lambda_{32}\Phi_{23}$.

And the last term of eq.(16) on the right hand side, the nonlinear term, is:

$$\left(\boldsymbol{\kappa}\tilde{\boldsymbol{\eta}}_0\right)_\bullet = \kappa\Phi_{2\bullet} \sum \boldsymbol{\Theta} + \mathrm{CC}, \tag{A.5}$$

where $\boldsymbol{\Theta}$ is comprised of many terms with different frequencies, up to 90 if both forces $f_1$ and $f_2$ are considered to have two distinct frequencies. These terms are omitted for brevity.